\documentclass[%
 reprint,
 amsmath,amssymb,
 aps,
]{revtex4-2}

\usepackage{graphicx}% Include figure files
\usepackage{dcolumn}% Align table columns on decimal point
\usepackage{bm}% bold math
\usepackage{color}% bold math
 \UseRawInputEncoding
\begin{document}

\preprint{APS/123-QED}

\title{Viewing explosion models of type Ia supernovae \\ via the insight from terrestrial cellular detonation}% Force line breaks with \\
%\title{Manuscript Title:\\with Forced Linebreak}% Force line breaks with \\
\thanks{A footnote to the article title}%

\author{Kazuya Iwata}
%% \altaffiliation[Also at ]{Physics Department, XYZ University.}%Lines break automatically or can be forced with \\
 \email{iwata.kazuya.6f@kyoto-u.ac.jp}
\affiliation{%
 Department of Mechanical Engineering and Science, Kyoto University\\
 Kyotodaigaku-Katsura, Nishikyo-ku, Kyoto, Japan 
}%

\author{Keiichi Maeda}%
\affiliation{
 Department of Astronomy, Kyoto University\\
 Kitashirakawa-Oiwake-cho, Sakyo-ku, Kyoto, Japan% with \\
}%

%%\collaboration{CLEO Collaboration}%\noaffiliation

\date{\today}% It is always \today, today,

\begin{abstract}
The cellular structure is considered to be a key as a criterion in \textcolor{black}{initiation}, propagation, and quenching of terrestrial detonation. \textcolor{black}{While a few studies on type Ia supernovae, which are known to involve detonation, have addressed the importance of the cellular structure, further detailed treatment will benefit enhanced understanding of the explosion outcomes.} In the present study, we bridge this gap in the astrophysics and engineering fields, focusing on the detonation in a helium-rich white dwarf envelope as the triggering process for the so-called double-detonation model. The cellular structures are quantified via high-resolution two-dimensional simulations. We demonstrate that widely-accepted terrestrial-experimental criteria for quenching and initiation of detonation can indeed explain the results of previous hydrodynamic simulations very well. The present study highlights the potential of \textcolor{black}{continuing to apply} the insight from terrestrial detonation experiments to astrophysical problems, specifically the \textcolor{black}{long} unresolved problem of the explosion mechanism of type Ia supernovae. 
%\begin{description}
%\end{description}
\end{abstract}

\maketitle

%\tableofcontents

\begin{figure}[t]
\centering
%%\begin{center}
\includegraphics[width=245pt,trim={140 10 180 5},clip]{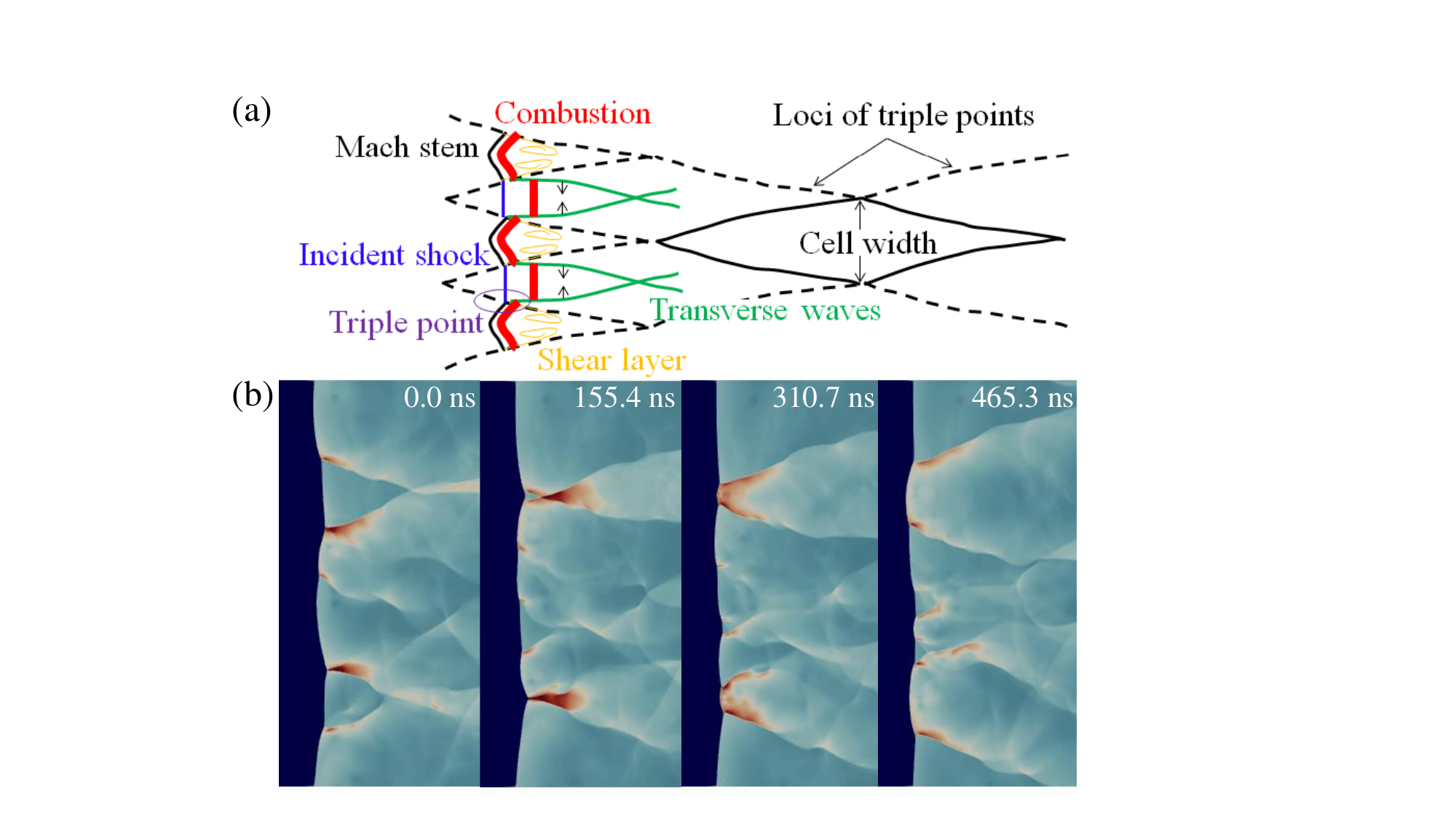}
%%\scalebox{0.6}{\includegraphics[trim={10 10 20 5},clip]{Figshematicanimation.pdf}}
\caption{(a) A schematic view of cellular structure of detonation, and (b) sequential images of cellular detonation observed in our simulations, with He:C:O=0.6:0.2:0.2 in mass fractions and $1\times10^6$ g cm$^{-3}$ in density. \label{fig:schematic}}
\end{figure}

%\section{\label{sec:level1}Introduction}
\textit{Introduction}---In spite of the major importance of type Ia supernovae (SNe Ia) as the standardized candle to measure the cosmological distance \cite{Phillips_1993}, details in their progenitor system(s) and explosion mechanism(s) have not been clarified, forming an important open issue \cite{Branch_1995, Maeda2016}. It is generally accepted that detonation must be involved, which is a supersonic burning front, propagating in and/or around a mass-accreting white dwarf (WD). This astrophysical detonation has a close analogy to terrestrial detonation observed in coal-mine explosions and detonation-powered engines \cite{lee_2008,WOLANSKI2013125}. Differences from terrestrial detonation originate from the \textcolor{black}{high-energy density} of astrophysical detonation: namely, electron degeneracy, energy and pressure of radiation, the multi-stage nature of nuclear burning as a powering source, \textcolor{black}{the extreme temperature sensitivity of nuclear reactions,} and so on. Nevertheless, they share a key common underlying physics - cellular structure, in which multi-dimensional bifurcated shock complex appears due to the intrinsic instability of the exothermic shock front as illustrated schematically in Fig.\ref{fig:schematic}(a). The development of the cellular structure has been demonstrated by a few works for the carbon-detonation inside a WD \cite{Boisseau_1996,Gamezo_1999,Timmes_2000,Papatheodore_2014}, and by our simulations for the helium-detonation in the helium-rich envelope (Fig.\ref{fig:schematic}(b)). \textcolor{black}{One key previous study on helium-detonation is Moore et al.\cite{Moore_2013}, who demonstrated development of the cellular structure, and discussed the criteria for detonation propagation and quenching. These works indicate} that the dynamical behaviors of terrestrial and astrophysical detonations have much in common. However, outcome of the cellular physics on astrophysical detonation has not been discussed in details. \\
It should be worthwhile considering the potential use of the theories and findings established in terrestrial detonation to astrophysical detonation problems. In fact, the criteria of \textcolor{black}{initiation}/propagation/quenching of detonation have been extensively studied in terrestrial chemical systems. These fundamental understandings are so established that they are commonly applied to design the detonation engines and to model detonation fronts in them. Most of these understandings are based on the spatial scale of cellular detonation, the so-called `cell width'. However, the application of the cell width and related cellular physics has been largely missing in studies of the SN Ia explosion mechanism.\\
In this study, helium-rich detonation is mainly addressed. It is the first stage of the double detonation model in the context of a sub-Chandrasekhar mass WD explosion, which is enjoying the revival for its availability to the double-degenerate (DD) channel \cite{IbenTutukov_1984,Webbink_1984, Shen_2018}. In the double-detonation model, first detonation initiated at the helium-rich WD envelope subsequently triggers secondary detonation in the carboy/oxygen core, leading to disruption of the entire WD. Cellular structure has not been investigated in such systems except for \cite{Moore_2013}, in which the cellular structure was seen in their simulations. In the present work, previous hydrodynamic simulations on the criteria of success and failure of helium-rich detonation \cite{Moore_2013,Shen_2014} are revisited from the perspective of the cellular physics. 

\textit{Numerical setup}---Considering the inherent invariance of cellular behavior in each lateral dimension, 2D cellular structure is addressed presently, as commonly done in simulations of terrestrial detonation. \textcolor{black}{This 2D treatment captures lateral transverse modes properly, since triagonal or spinning modes are not important in the problem under consideration.} Our in-house 2D hydrodynamic code, which has been used to simulate a wide range of terrestrial detonation problems \cite{Iwata_2016,Iwata_2023}, has been modified to include nuclear reaction network and equation of state (EOS) for high-density and fully-ionized plasma. Most of its numerical methodologies are common to those employed in our previous one-dimensional study \cite{Iwata_2022}. It uses an Eulerian method to solve reactive Euler equation systems. \textcolor{black}{We note that the heat conduction is negligible in the detonation problem considered here, unlike the case for deflagration.} The continuity equations of 13 isotopes are included, which are connected through an alpha-chain nuclear reaction network. \textcolor{black}{Hydrogen burning and photo-disintegration included in the 19-isotope network in \cite{nineteen_network} are not considered.} We defer investigation of the boosting effect of the proton-catalyzed reaction sequence ($^{12}\rm{C}(\it{p},\gamma)^{\rm{13}}\rm{N}(\alpha,\it{p})^{\rm{16}}\rm{O}$) 
\cite{Shen_2014} to our future work, in order to have the same network with the previous studies for fair comparison. Gravity is not included in the equations. Timmes EOS \cite{Timmes_1999} is used in a tabulated form to obtain temperature and pressure from internal energy.\\
Our 2D model setup shares much in common with that in \cite{Gamezo_1999}. It is intended to acquire steady Chapman-Jouguet (C-J) detonation, the cell width of which is used as a basis of the detonation criteria in terrestrial systems \cite{Chapman1899,Detonation_database}. The computational domain is rectangular, divided by equally-spaced square meshes. A detonation-fixed coordinate frame is used to capture the steady propagation with a fewer computational cost, in which inertial force is added to control the averaged longitudinal position of detonation within a few meshes. The fixed inflow is imposed on the left boundary and the flow goes out through the right boundary. \textcolor{black}{To ensure the stable propagation at the C-J velocity, the distribution of the meshes in the downstream region is extended longitudinally far enough to reach the equilibrium state calculated via the 1D Zeldovich-von-Neumann-Doring (ZND) theory (\cite{Zeldovich1940,Neuman1942TheoryOD,Doering}).} The lateral (upper and lower) boundaries are treated as periodic, which is different from \cite{Gamezo_1999} who applied reflective boundaries; \cite{Boisseau_1996} indicated that the choice would not affect the cellular structure. Indeed, we confirmed that extending the lateral domain size has insignificant influence on the observed cell width.\\
The upstream density is varied to be $10^5$, $2\times10^5$, $5\times10^5$ and $10^6$ g cm$^{-3}$, to represent the condition at the base of the WD envelope with core masses of 0.9-1.1 M$_\odot$ and He envelope masses of 0.01-0.05 M$_\odot$. The mass fraction of ${^4}\rm{He}$ ($X_{\rm{He}}$) is also varied to be 0.0, 0.6, 0.8, 0.9 and 1.0, with the rest divided equally in masses to $^{12}\rm{C}$ and $^{\rm{16}}\rm{O}$. Thus, twenty cases in total are simulated. The $X_{\rm{He}}$=0.0 cases are for comparison to the conditions of secondary carbon/oxygen detonation in the WD core material. A resolution for each computational case is chosen so that twenty meshes are placed within the energy release scale $L_{\rm{q}}$, which is defined based on the 1D ZND theory as the post-shock distance at which the total accumulated amount of nuclear energy release  (with the rate described by $\dot{\epsilon}_{nuc}$) becomes comparable to the initial internal energy $\epsilon_{int}$:
\begin{equation}\label{eq:Lq}
\epsilon_{int} = C\int^{L_q}_0 \dot{\epsilon}_{nuc}dx/u \ ,
\end{equation}
where $x$ and $u$ are distance from the shock and 1D velocity, respectively. $C$ is set to 0.20 so that $L_q$ is almost equal to the half-reaction length of carbon fuel in the $X_{\rm{He}}$=0.0 cases. We confirmed that this resolution level was enough for mesh convergence \textcolor{black}{within a factor of 1.6. This is reasonable even in the state-of-the-art simulations of terrestrial detonation, which tends to suffer from irregular variation by a few factor} \cite{Mazaheri2013}. 
%The upstream temperature, $T_\rm{1}$, is set to be constant at $T_1$=$2\times10^8$ K; 
%the choice of $T_\rm{1}$ is confirmed to hardly affect the structure of the detonations. 

%%\section{2D Cellular structure} \label{sec:style}

\begin{figure}[t]
\centering
\includegraphics[width=246pt,trim={1.5 1 240 1},clip]{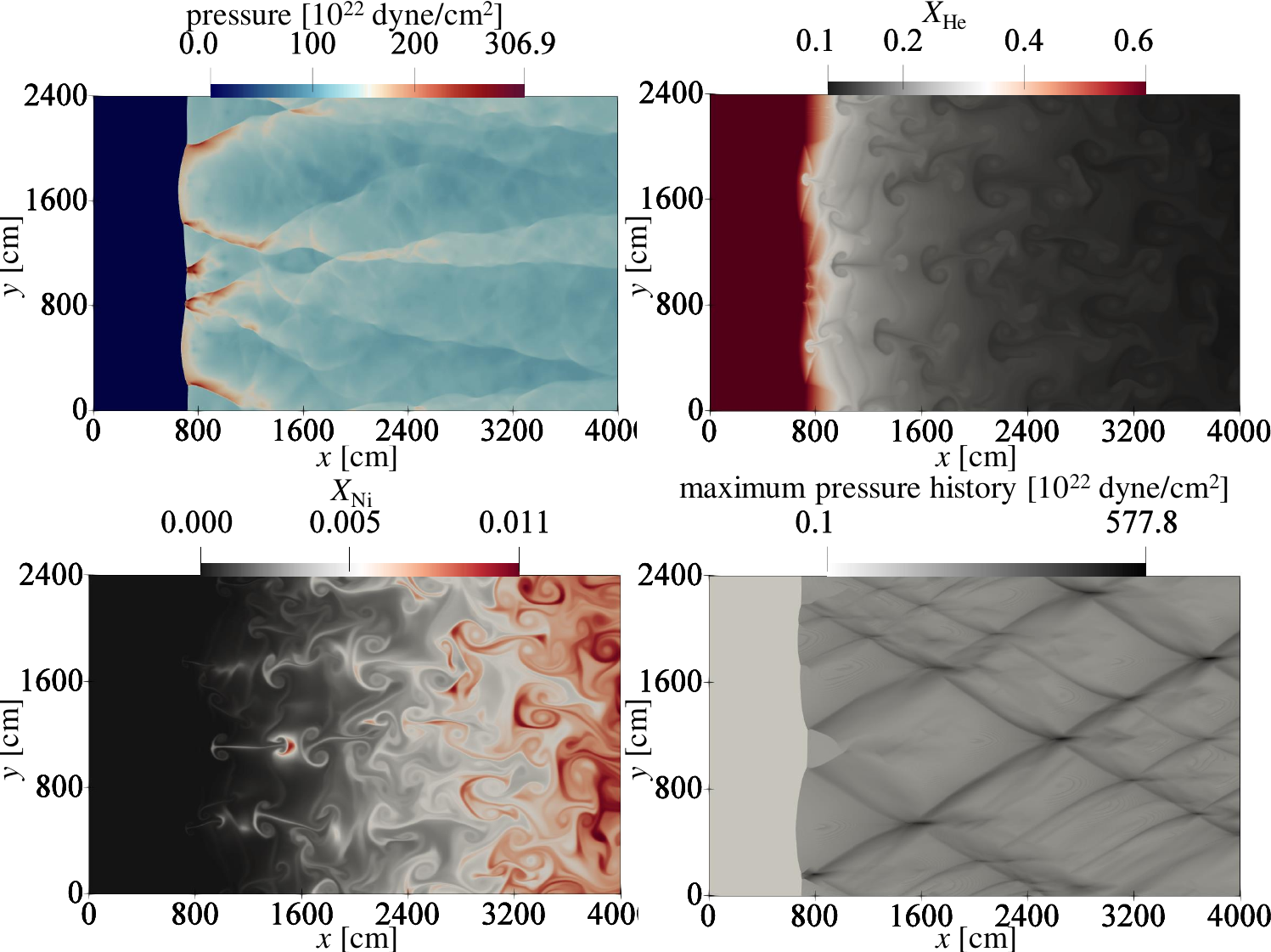}
%%\scalebox{0.7}{\includegraphics[trim={77 60 10 1},clip]{Figcell1.pdf}}
\caption{Cellular structure in the $X_{\rm{He}}$=0.6 and $1\times10^6$ g cm$^{-3}$ case, shown as the distributions of pressure, mass fractions of $^{\rm{4}}\rm{He}$ and $^{\rm{56}}\rm{Ni}$, and the maximum pressure history. Coordinates are shown in cm. \label{fig:pressuremaxp}}
\end{figure}

\begin{figure}[t]
\centering
\includegraphics[width=246pt,trim={1.5 1 240 1},clip]{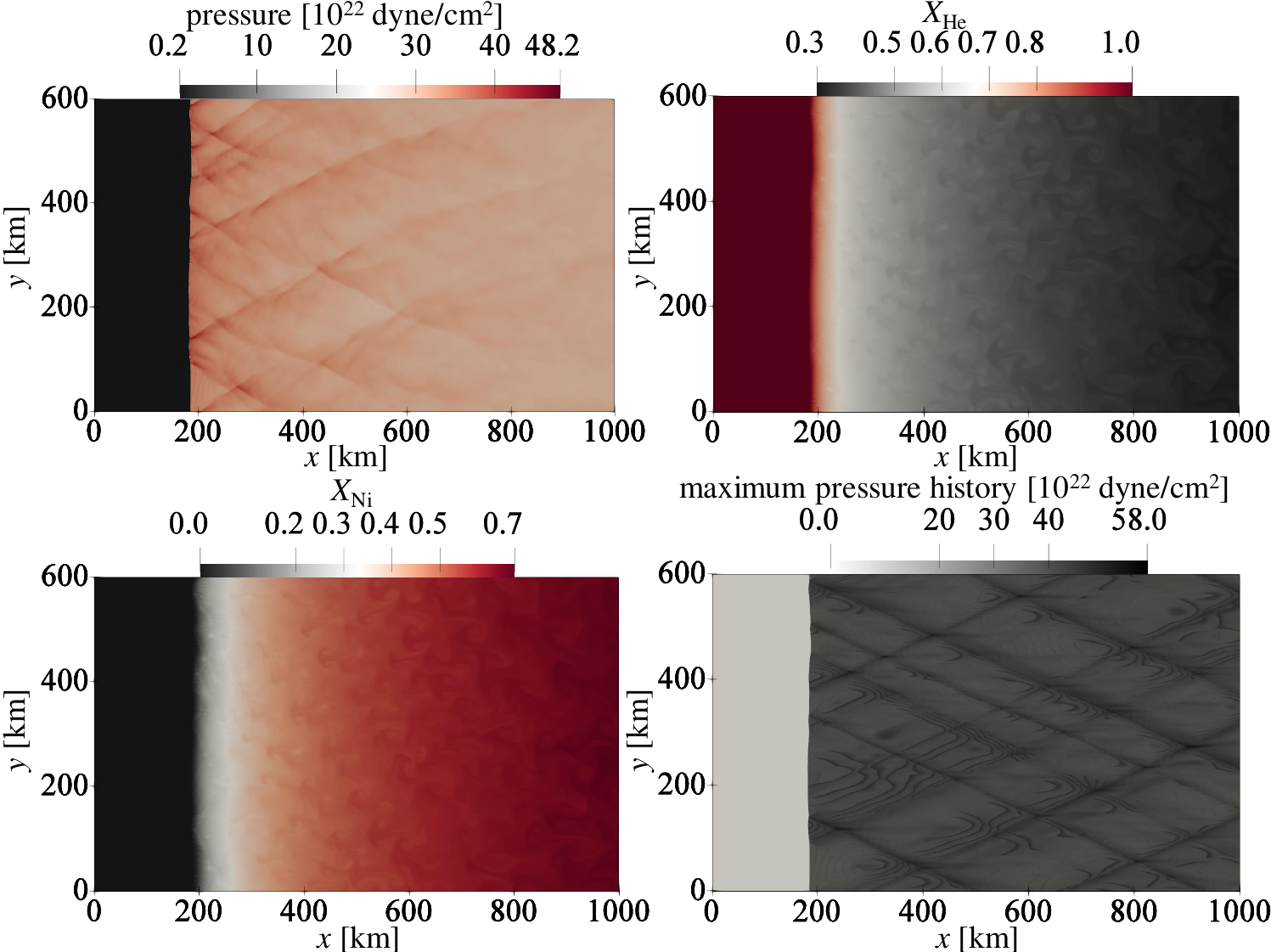}
%%\scalebox{0.7}{\includegraphics[trim={77 60 10 1},clip]{Figcell2.pdf}}
\caption{Cellular structure in the $X_{\rm{He}}$=1.0 and $2\times10^5$ g cm$^{-3}$ case (see the Fig. \ref{fig:pressuremaxp} caption). Coordinates are shown in km. \label{fig:pressuremaxp2}}
\end{figure}

%\section{\label{sec:level1}2D Cellular structure}
\textit{2D Cellular structure}---Two computational cases, one with $X_{\rm{He}}$=0.6 and $1\times10^6$ g cm$^{-3}$, and the other with $X_{\rm{He}}$=1.0 and $2\times10^5$ g cm$^{-3}$, are selected for demonstrating the cellular structures of the helium detonation; Figures \ref{fig:pressuremaxp} and \ref{fig:pressuremaxp2} show snap-shot distributions of pressure as well as the mass fractions of $^{\rm{4}}\rm{He}$ and $^{\rm{56}}\rm{Ni}$, along with the maximum pressure histories experienced at each computational mesh as a numerical soot-foil record for cell width. The mesh size is 2.0 cm and 0.50 km for these cases, respectively. A peaky-pressure distribution is evident for the $X_{\rm{He}}$=0.6 and $1\times10^6$ g cm$^{-3}$ case with strong vortical motions that appear behind the bifurcated shock waves. Pressure is more uniform throughout the domain in the $X_{\rm{He}}$=1.0 and $2\times10^5$ g cm$^{-3}$ case, approaching a planar shock-flame front. Furthermore, the largest difference to be highlighted is the size of the cellular structures: $8\times10^2$ cm and $1\times10^2$ km, respectively. \\

\begin{figure}[t]
\centering
\includegraphics[width=305pt,trim={300 305 205 12},clip]{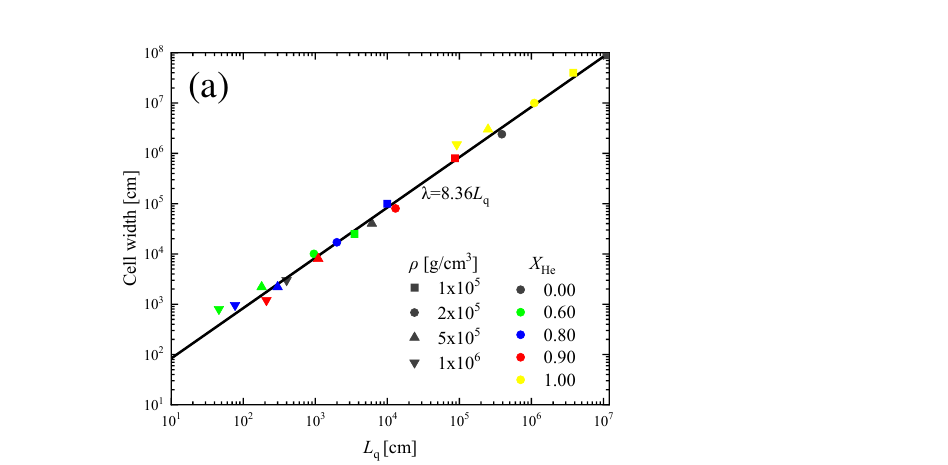}
\includegraphics[width=305pt,trim={185 90 343 230},clip]{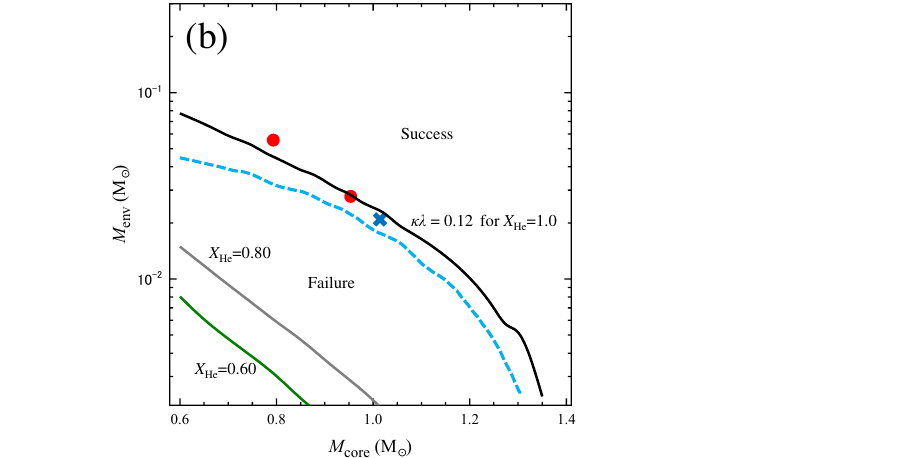}
\includegraphics[width=305pt,trim={428 93 100 230},clip]{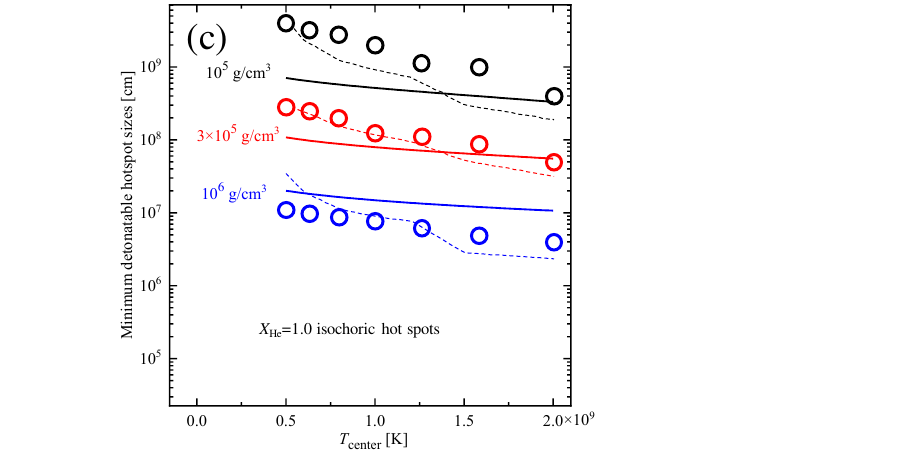}
%%\scalebox{0.7}{\includegraphics[trim={125 76 63 12},clip]{Figanalysis.pdf}}
\caption{(a) Cell widths versus energy release scale $L_q$, (b) success and failure of detonation propagation in the WD envelope in the $M_{core}$-$M_{env}$ diagram, and (c) the minimum isochoric hotspot size needed for initiating pure He detonation shown against the central temperature $T_{center}$.  \label{fig:cellwidth}}
\end{figure}

The cell widths are derived from the number of transverse waves observed in each numerical soot foil, and plotted as symbols in Fig.\ref{fig:cellwidth}(a), where different-color symbols represent different $X_{\rm{He}}$. %A range of cell width is observed, depending on the density and composition (e.g., Figs. \ref{fig:pressuremaxp} and \ref{fig:pressuremaxp2}). Despite the large range range found for the cell width, 
The cell width shows a good linear correlation with the energy release scale $L_{\rm{q}}$, which reflects the underlying physics of exothermicity-driven shock instabilities. A black solid line denotes the linear relationship derived by the least-square fit, which results in 
%\begin{equation}\label{eq:lambdaLq}
% \epsilon_{int} = C\int^{L_q}_0 \dot{\epsilon}_{nuc}dx/u 
%\end{equation}
\begin{equation}\label{eq:lambdaLq}
\lambda = 8.36 L_{q} \ .
\end{equation}

%%\section{Estimate of success and failure of detonation} \label{sec:style}
%%\section{\label{sec:level1}Estimate of success and failure of detonation}
{\textit{Criterion for the detonation propagation/quenching}---\textcolor{black}{In the actual circumstances in stellar media, detonation can be deformed into a curved shape owing to the gradients of density, composition and other inhomogeneities, deviating from C-J detonation discussed in the previous section. The curvature will delay the post-shock reactions by expansion wave, and could lead to detonation quenching if/when the effect is sufficiently strong.} According to \cite{Nakayama2013} who empirically determined the threshold of the curvature-induced quenching based on their experimental results in a terrestrial-chemical system, detonation ceases to propagate when the wave curvature $\kappa$ and the cell width \textcolor{black}{at the planar C-J state} $\lambda$ satisfy the relation $\kappa\lambda$ $\gtrsim$ 0.12. In addition, \cite{Moore_2013} simulated He detonation at the base of a WD envelope, and their simulation results indicate the relationship $1/\kappa=1.5H_{\rm{p}}$, where $H_{\rm{p}}$ is pressure scale height determined by the hydrostatic structure of a WD. By combining these relations, we derive the quenching threshold as follows;\\
\begin{equation}\label{eq:quenching}
\lambda/H_{p} = 0.18 \ .
\end{equation}
In \cite{Moore_2013}, the formulation of generalized 1D ZND theory was constructed including the effects of curvature and expansion, with which the threshold for the success and failure of pure He detonation in the $\it{M}_{\rm{core}}$ - $\it{M}_{\rm{env}}$ diagram was derived (light-blue dashed curve in Fig.\ref{fig:cellwidth}(b)). It was derived as \textcolor{black}{the thinnest He envelope for which a propagating solution of the modified ZND model is found, which is consistent with} the condition for which $^{40}$Ca becomes the dominant product. In the same figure, circles and a cross are plotted as their simulation results, denoting the success and failure of pure He detonation, respectively. Our prediction based on Eq. \ref{eq:quenching} is illustrated as a black solid curve, in which $\lambda$ for the corresponding $\it{M}_{\rm{core}}$ and $\it{M}_{\rm{env}}$ is calculated using Eq. \ref{eq:lambdaLq} (as derived by our local-scale simulations) based on 1D ZND consideration. 
\textcolor{black}{It is seen that our threshold, as constructed by combining the insight from terrestrial experiments and the outcomes from hydrodynamic study of \cite{Moore_2013}, provides an improved criterion on the success and failure}. Even though more data from large-scale simulations will be needed to provide a statistically robust validation, it demonstrates \textcolor{black}{that the application of the terrestrial-experiment-based insight into analysis on the outcome of SNe Ia is promising}. Further advantage of using this terrestrial threshold is its general versatility; \textcolor{black}{a rough estimate of} the explosion outcome \textcolor{black}{is possible with the assistance of the related hydrodynamic studies, thereby reducing efforts spent on} large-scale hydrodynamic simulations, and we can also apply it to different explosion models (e.g., delayed-detonation, the pulsating detonation).\\
Mixing of the carbon/oxygen core material into the envelope highlights another important implication; as represented by gray and green curves in Fig. \ref{fig:cellwidth}(b), the core-envelope mixing substantially reduces the required \textcolor{black}{He} envelope mass $M_{env}$ for the successful surface-detonation propagation. In this case, detonation is feasible with $M_{env}$ well below 0.02 $\rm{M}_\odot$, which is the observationally-derived upper limit for the double-detonation model in order to explain properties of normal SNe Ia \cite{Woosley_2011,Maeda_2018,Shen_2021}. This is in accordance with \cite{Shen_2014}, who attributed this decreasing behavior to the enhanced energy generation via the $\alpha$-capture process of carbon as $^{12}$C($\alpha$,$\gamma$)$^{16}$O. We here reinterpret this trend in terms of the terrestrial-experimental insight; this far milder requirement is attributed to smaller cell width with intermediate $^{4}$He mass fractions as shown in Fig.\ref{fig:cellwidth}(a). \textcolor{black}{Note that the discussion based on \textcolor{black}{the} ZND theory \cite{Shen_2014} reached a similar conclusion; indeed, two approaches are bridged through Eq.\ref{eq:lambdaLq}.}\\
As represented in Fig.\ref{fig:cellwidth}(a), the minimum cell width reaches the order of 10$^2$ cm for the density of 10$^6 \rm{g cm}^{-3}$, \textcolor{black}{in case of the core-envelope mixed conditions.} Such a small scale is extremely difficult to resolve in full-star studies; actually, the mesh size has been limited to a few km even in the highest resolution cases found in previous large-scale (full-star) simulations \cite{Pakmor_2022,Boos2021,Guillochon_2010}. Our present method based on the terrestrial-experimental insight, \textcolor{black}{as coupled with 1D-ZND approaches like done in \cite{Moore_2013,Shen_2014}}, can address the progenitor conditions where \textcolor{black}{resolved} full-star simulations are \textcolor{black}{challenging, even though we need to compare our methods with future highly-resolved studies in the mixed He envelopes. As a highlight, the importance of the core-envelope mixing in the double-detonation model for explaining normal SNe Ia, which has been assessed so far relying on hydrodynamic approaches\cite{Shen_2014,Moore_2013,Boos2021,Townsley_2019}, is demonstrated by a new insight from terrestrial detonation experiments.} 

\textit{Criterion for the detonation \textcolor{black}{initiation}}---Another application is the minimum energy required to initiate detonation. In many previous hydrodynamic studies on the double-detonation model, an energetic hotspot is artificially inserted to force the \textcolor{black}{initiation} of He detonation on a WD surface (\cite{Boos2021,Fink2010,Gronow2020SNeIF}). However, according to \cite{Iwata_2022}, detonation is very hard to directly ignite in the WD envelope except for an extremely heavy progenitor and/or a massive \textcolor{black}{He} envelope: a burning front is most likely to start as deflagration, which is a subsonic flame. In \textcolor{black}{spherical 1D} simulations of \cite{Shen_2014}, the hostspot size and its central temperature were varied to investigate the minimum hotspot size required for detonation initiation, and the required size was found to be beyond the WD radius for low-mass \textcolor{black}{He} envelopes. \textcolor{black}{They considered the gradient ignition mechanism, which corresponds to the so-called SWACER mechanism in terrestrial detonation\cite{LEE1980359}, as the key mechanism of detonation initiation. In the SWACER mechanism, deflagration-to-detonation transition (DDT) can occur when the spatial gradient of induction time is close to the inverse of C-J velocity; a resonant acceleration to a supersonic regime is then triggered. Our approach considering cellular physics provides a complementary view and enhance understanding of detonation initiation, since multidimensional aspect of cellular structure is intrinsic for stabilizing detonation through the local explosion occurring at the collision of transverse waves (as demonstrated in Fig. \ref{fig:schematic}(b) and overviewed in \cite{Lee_surface_1984,lee_2008}).}\\ 
We address the topic of detonation initiation in view of the condition for \textcolor{black}{the direct initiation of detonation} established based on terrestrial experiments. For terrestrial detonation, the requirement for directly initiating detonation was proposed by \cite{lee_2008,Lee_surface_1984} as the surface energy theory. The formulation for this theory is the following:
\begin{equation}\label{eq:surface}
E_{\rm{min}} = \frac{2197}{16}\pi\rho_{\rm{ini}}I_{\rm{1}}D^2_{\rm{CJ}}\lambda^3 \ ,
%D/D_{CJ} = 1-1.3017\kappa\lambda+16.089(\kappa\lambda)^2-169.67(\kappa\lambda)^3
\end{equation}
where $I_{\rm{1}}$=0.423 (for the specific heat ratio of $\gamma$=1.4) \textcolor{black}{was used in \cite{Lee_surface_1984} to compare with measured $E_{\rm{min}}$ for ideal-gas mixtures}. \textcolor{black}{For the present mixture condition, $\gamma$ is around 1.6, and it will lead to the variation of $I_{\rm{1}}$ within 0.4-1.0. However, it will translate to the errors of $E_{\rm{min}}$ only within a factor of 1.36, and thus this effect is not important for the present purpose.} $\rho_{\rm{ini}}$ and $D_{\rm{CJ}}$ are initial density and the C-J velocity (i.e., the theoretically-predicted propagation velocity of detonation \cite{Chapman1899}), respectively. A cubic dependence on cell width $\lambda$ makes the behavior of this minimum energy dominated by cell width. This formulation is known to provide good prediction of experimentally observed minimum \textcolor{black}{initiation} energy $E_{\rm{min}}$ \textcolor{black}{with $\gamma$ =1.4} for a wide variety of fuels \cite{Lee_surface_1984}. In the present work, we use Eq. \ref{eq:surface} to estimate the requirement of the hotspot size by evaluating the internal energy contained in the spot. A uniform density and linear distribution of temperature ($\it{T}=\it{T}_{\rm{center}}$ at the center and with $\it{T}=\rm{10}^7$ K at the edge) are assumed within the hotspot, i.e., isochoric hostspots \cite{Shen_2014}. Cell width \textcolor{black}{at the planar C-J state} $\lambda$ is calculated using Eq. \ref{eq:lambdaLq}. \textcolor{black}{Note that we only consider an isochoric hot spot, since local pressure is built up quickly relative to sound crossing time in an actual condition feasible for direct detonation initiation.} \\
Fig. \ref{fig:cellwidth}(c) compares our analytical results (solid curves) and 1D simulations of \cite{Shen_2014} \textcolor{black}{(dotted lines) and Holcomb et al. \cite{Holcomb_2013}}. In the latter, the minimum hotspot size for the successful detonation was determined iteratively by varying its radius at each value of $\it{T}_{\rm{center}}$. A rough match is found for the three different initial densities. Discrepancies are larger for lower $\it{T}_{\rm{center}}$ and lower initial density, but are within a factor of a few. Therefore, Eq. \ref{eq:surface} is confirmed to provide a rough but appropriate estimate for the direct occurrence of detonation. \textcolor{black}{An} estimation based on terrestrial experiments can thus provide useful and quick diagnostic to quantify the progenitor conditions likely to trigger the double-detonation. 

\textit{Summary}---\textcolor{black}{In view of} the analogies in fundamental physics between astrophysical detonation and terrestrial detonation, \textcolor{black}{further enhancing interaction between the two fields is important for deepening our understanding of detonating mechanisms in astrophysical objects.} In the present work, the potential of laboratory chemical-detonation experiments was demonstrated to \textcolor{black}{inform the detailed treatment of astrophysical detonations}, presently focusing on the first stage of the double-detonation model. The criteria for \textcolor{black}{initiation} and propagation/quenching of the detonation within the WD envelope agreed very well with previous hydrodynamic simulations. Our method can also provide insights into the progenitor conditions that require to resolve the small scale beyond current computational resources can handle in full-star simulations. These findings benefit enhanced understanding of open problems of the explosion mechanism of SNe Ia via simple analytical tools that have been calibrated with reliable experimental laws. %The present work highlights a power of introducing insights from terrestrial combustion experiments to the astrophysical problem, especially on the detonation physics.\\
%We have to make continuing efforts to quantify the outcome of the explosion models of SNe Ia from the perspective of cellular dynamics and related terrestrial combustion physics.  \\

\begin{acknowledgments}
The numerical calculations were carried out on Yukawa-21 at YITP at Kyoto University and on Cray XC50 at Center for Computational Astrophysics at National Astronomical Observatory of Japan. K. I. acknowledges support from the Japan Society for the Promotion of Science (JSPS) KAKENHI grant 23K13146. K.M. acknowledges support from JSPS KAKENHI grant (JP18H05223, JP20H00174, and JP24H01810) and support from The Kyoto University Foundation. 
\end{acknowledgments}

\bibliography{arxiv}% Produces the bibliography via BibTeX.

\end{document}